# EXTRATERRESTRIAL LIFE AND CENSORSHIP

N. Chandra Wickramasinghe

Cardiff Centre for Astrobiology, Llwynypia Road, Cardiff CF14 0SY, UK

E mail: ncwick@gmail.com

**Abstract**

In this article I chronicle a series of landmark events, with which I was personally involved, that relate to the development of the theory of cosmic life. The interpretation of events offered here might invite a sense of incredulity on the part of the reader, but the facts themselves are unimpeachable in regard to their authenticity. Of particular interest are accounts of interactions between key players in an unfolding drama connected with the origins of life. Attempts to censor evidence incompatible with the cosmic life theory are beginning to look futile and a long-overdue paradigm shift may have to be conceded.

1. **Introduction**

The ingress of alien microbial life onto our planet, whether dead or alive should not by any rational argument be perceived as a cause for concern.  This is particularly so if, as appears likely, a similar process of microbial injection has continued throughout geological time.  Unlike the prospect of discovering alien intelligence which might be justifiably viewed with apprehension, the humblest of microbial life-forms occurring extraterrestrially would not constitute a threat.  Neither would the discovery of alien microbes impinge on any issues of national sovereignty or defence, nor challenge our cherished position as the dominant life-form in our corner of the Universe.

Over the past three decades we have witnessed a rapid growth of evidence for extraterrestrial microbial life.  Along with it has grown a tendency on the part of scientific establishments to deny or denounce the data or even denigrate the advocates of alien life.  My own personal involvement in this matter dates back to the 1970's when, together with the late Fred Hoyle, I was investigating the nature of interstellar dust.  At this time evidence for organic molecules in interstellar clouds was accumulating at a rapid pace, and the interstellar dust grains that were hitherto believed to be comprised of inorganic ices were shown by us to contain complex organic polymers of possible biological provenance (Wickramasinghe, 1974; Wickramasinghe, et al, 1977; Hoyle and Wickramasinghe, 1977a,b).  These discoveries came as a surprise to astronomers, and for a long time the conclusion was resisted that such molecules might have a relevance to life on the Earth (Hoyle and Wickramasinghe, 1986a,b).

Biologists in the 1960's and 1970's had no inkling of the intimate connection of their subject with astronomy.  The holy grail of biology was the hypothesis that life emerged from a primordial soup generated *in situ* from inorganic molecules on the primitive Earth (Oparin, 1953).   The primordial soup theory gained empirical support from the classic studies of Miller (1953) and Miller and Urey (1959) that showed the production of minute quantities of amino acids and sugars by sparking mixtures of inorganic gases.  Ponnamperuma and Mack (1965) later demonstrated the production of nucleotides (components of DNA) under similar conditions in the laboratory.  Finally the experiments of Sagan and Khare (1971) showed the production of amino acids from gases exposed to ultraviolet light (1971).  All such experimental triumphs were greeted as crucial steps towards understanding the origin of life on the Earth, although it was never clear that the experimental conditions used in the laboratory had any relevance to the primitive Earth. The early terrestrial atmosphere that is now believed to have been oxidising would have inhibited any synthesis of organics of the type demonstrated by Miller, Sagan and Ponnamperuma.

The hypothesis of the terrestrial origin of the chemical building blocks of life might have been thought plausible before it was discovered that vast quantities of biogenic organic molecules existed within the interstellar clouds (Hoyle et al, 1978; Kwok, 2009). Having first argued for a complex biochemical composition of interstellar dust, Hoyle and I were among the first to make a connection between complex organic molecules in interstellar clouds and life on Earth (Hoyle and Wickramasinghe, 1976, 1978, 1981). The total amount of organic material in the galaxy in the form of organic dust and PAH's accounts for about a third of all the carbon present in interstellar space – a truly vast quantity amounting to some billion or so solar masses (see review by Kwok, 2009).

2. **Censorship Dawns**

My first inkling of any censorship relating to extraterrestrial life came when we made the intellectual leap from prebiology in space to fully-fledged biology outside the Earth (Hoyle and Wickramasinghe, 1976, 1982; Hoyle et al, 1984). In setting out to explore the hypothesis that interstellar grains were not just abiotic organic polymers but bacterial cells in various stages of degradation, we made a prediction that interstellar dust in the infrared spectral region must have the signature of bacteria (Hoyle et al, 1982). Infrared sources near the galactic centre were a prime target for this investigation and on our instigation approaches were made to the Anglo-Australian telescope committees to provide time on the AAT to test our seemingly wild hypothesis. An application for observing time for this project made by my brother Dayal T. Wickramasinghe at ANU and David Allen was duly refused as "having no scientific value" (Wickramasinghe, 2005). In the event, Dayal, who was allocated time on the AAT for a totally different project, found a reason to illicitly study the spectrum of the Galactic Centre source GC-IRS7. When he did so he discovered an amazingly close fit to our predicted absorption curve for bacteria – a prediction that was made a full 3 months ahead of the serendipitous observations being made (Hoyle et al, 1982). See Fig. 1.

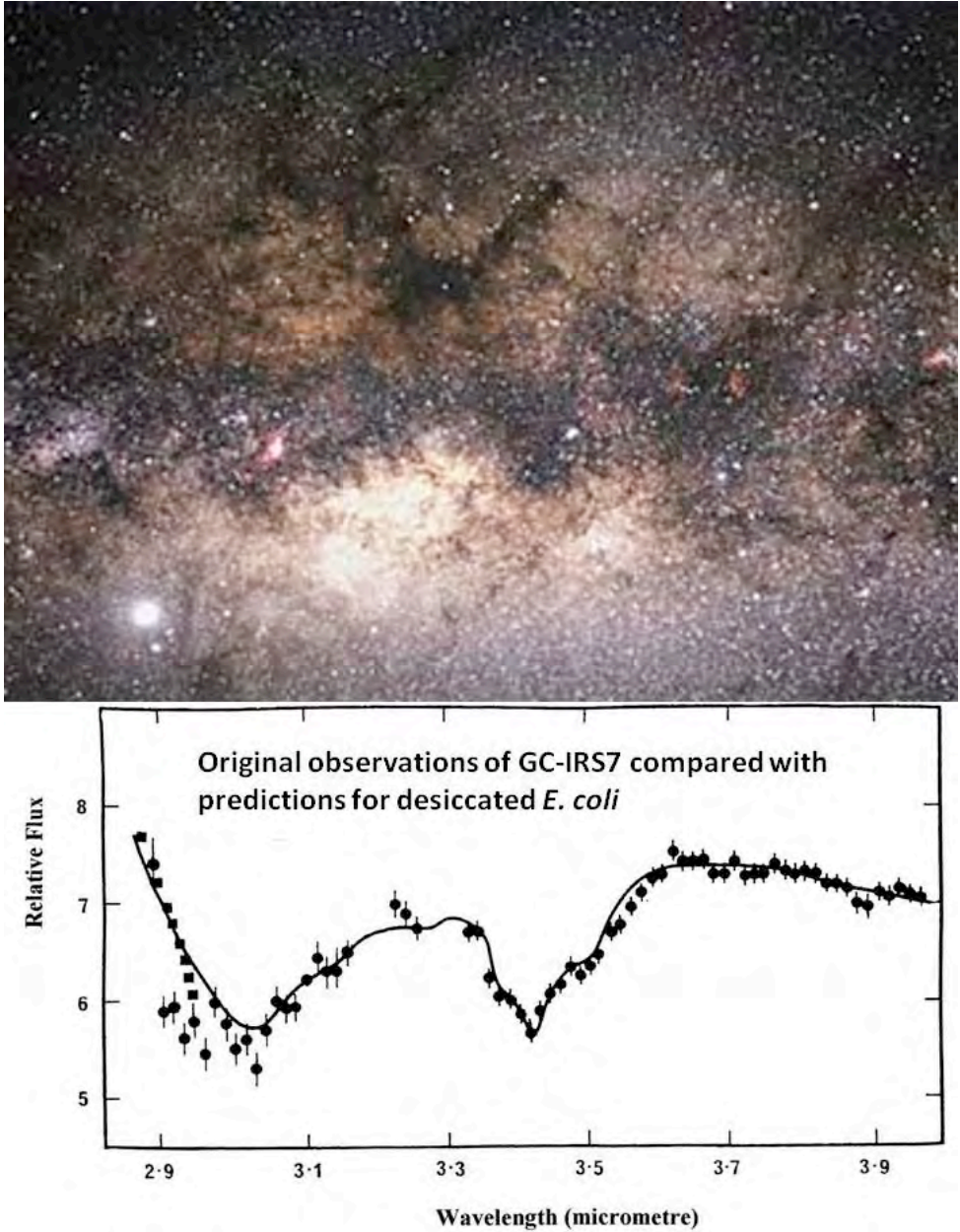

Fig. 1. The first observations by D.T. Wickramasinghe and D.A. Allen of the Galactic Centre source GC-IRS7 compared with the predictions for desiccated *E. Coli* (Hoyle, Wickramasinghe and Al-Mufti, 1984).

From 1982 to the present day astronomical observations of interstellar dust and molecules ranging in wavelength from the far ultraviolet to the infrared have continued to support the biological origin. It would appear that a large fraction of interstellar grains and molecules must have a biological provenance, implying that microbial life exists on a grand galactic or even cosmological scale (Wickramasinghe, 1994, 2010; Hoyle and Wickramasinghe, 2000,

J.Wickramasinghe et al, 2010; Gibson and Wickramasinghe, 2010; Gibson, Schild and Wickramasinghe, 2010).

After 1982, when evidence for cosmic life and panspermia acquired a status close to irrefutable, publication avenues that were hitherto readily available became suddenly closed. With the unexpected discovery that comets had an organic composition, with comet dust possessing infrared spectra consistent with biomaterial (Hoover et al, 1986; Hoyle and Wickramasinghe, 1986a,b) attitudes hardened to a point that panspermia and related issues were decreed taboo by all respectable journals and institutions.

The peer review system that was operated served not only to exclude poor quality research but also to deliberately filter publication of any work that challenged the standard theory of life's origins.

Even though the general public revelled in ideas of extraterrestrial life, science was expected to shun this subject no matter how strong the evidence, albeit through a conspiracy of silence. It was an unwritten doctrine of science that extraterrestrial life could not exist in our immediate vicinity, or, that if such life did exist, it could not have a connection with Earth.

3. **The Meteorite Microfossil Saga**

This campaign of explicit denials and censorship may have started between 1962 and 1965 when microorganisms were actually recovered from the stratosphere using balloons flown to heights between 20 and 43km. Although the lower heights in this range were not great enough to exclude terrestrial contamination, the density dependence with height of the recovered particles was consistent with an *infall* rate of $10^{19}$ cells per year (Greene et al, 1962-65). This important pioneering work, carried out by NASA at the dawn of the Space Age, probably rang alarm bells to which the authorities had to react, and react they surely did. I was told by Leslie Hale, an atmospheric scientist at Penn State University that this exciting programme of work was suddenly halted by funds being withdrawn. Nothing more was said.

Journals like *Nature* and *Science* had for a long time served as staunch guardians of reigning paradigms across the whole of science, straddling cosmology and biology, acting in a sense as the secular "protectors of the faith". Notwithstanding the constraints imposed by such a role these journals did in fact occasionally publish ground-breaking research that contradicted reigning paradigms. Sometimes they published highly controversial work and invited debate and refutation.

The first reports of the detection of microbial fossils in meteorites were published in the columns of *Nature* by Claus, Nagy and others (Claus and Nagy, 1961; Nagy et al, 1962, 1963). No sooner than these publications appeared a vigorous campaign of refutation and denial was mounted by Anders and others, also in the columns of *Nature* (Anders, 1962; Anders and Fitch. 1962; Fitch et al, 1962). With a ruthless and forceful denigration of these claims on grounds of alleged contamination, the microfossil saga faded from view for a full 20 years. And with a veritable army of nay-sayers braying so stridently the world became convinced that this was a quagmire to which one should never return.

The early pioneers of microfossil discovery were thus silenced and had little choice but to recant. I was told by a reliable witness that Claus was ruthlessly bullied into capitulation, and Nagy also retreated somewhat whilst continuing to hint in his writings that it might be so, rather in the manner of Galileo Galilei's whispered "*E pur si mouve*" – and yet it moves.

Whilst such rumblings continued the earlier work of Claus and Nagy came to be superseded in quality by investigations carried out by Hans Dieter Pflug in 1981. Twenty years on, with improved techniques of sample preparation, electron microscopy and laser ion probe spectroscopy, the signal-to-noise ratio for microfossil detection was improved by at least an order of magnitude. In 1980 Hans Pflug corresponded with Fred Hoyle and myself to inform us that his new studies corroborate our own by now well-publicised claims of extraterrestrial microbial life. In his investigation of the Murchison meteorite ultra-thin slices of the meteorite were placed on a perfectly clean membrane and the mineral matrix leached out with hydrofluoric acid, thus leaving any included organic structures intact. Pflug discovered a wide range of organic structures uncannily similar to terrestrial microorganisms; and with laser ion probe studies and EDAX analysis he found their chemical compositions and forms to be consistent with microbial fossils.

I invited Hans Pflug to visit us in Cardiff and on the 26th of November 1981 he delivered a lecture, introduced by Fred Hoyle, that left the audience speechless. The poster for the lecture and some of his key images are shown in Fig. 2. The response to Pflug's discoveries differed markedly from the earlier attitudes to Claus and Nagy. Pflug was not attacked on grounds of contamination or artifacts, but he was given what could be described as the "silent treatment". This was the same manner in which my own collaborations with Hoyle at the time were received. There were muffled whisperings in University common rooms, but nothing in the way of a well-formulated technical criticism ever appeared in print.

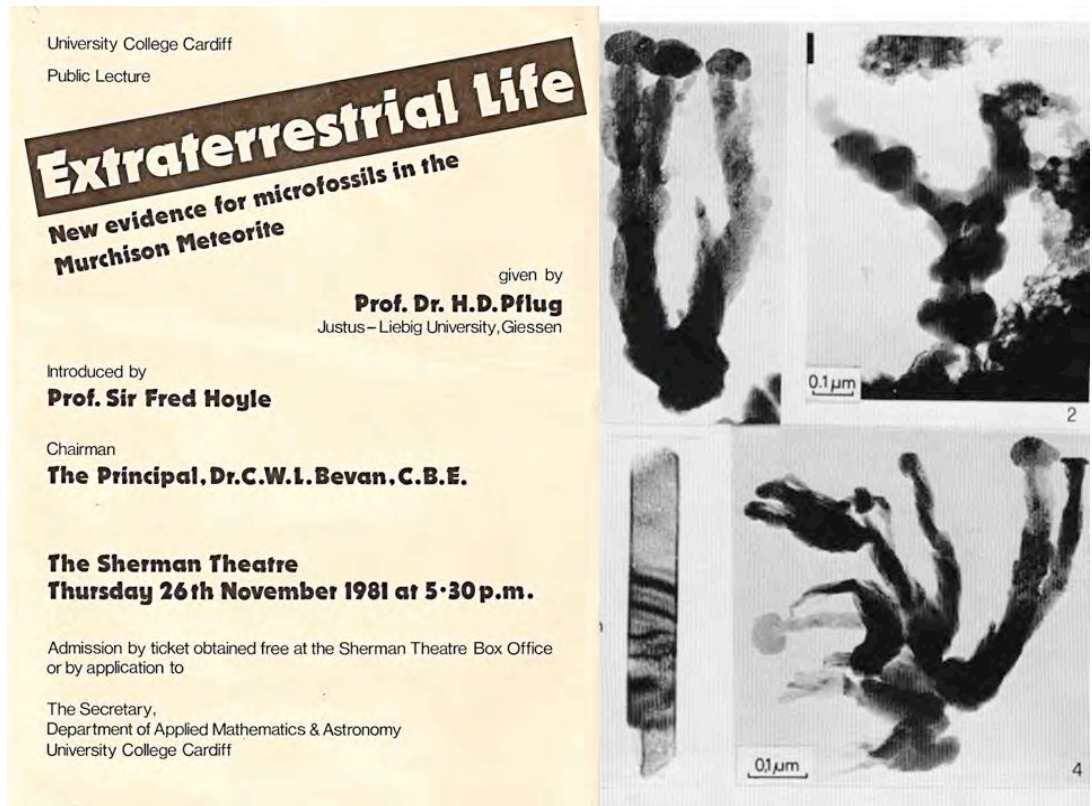

Fig. 2  Pflug's lecture poster and an example of a microfossil of *pedomicrobium* in the Murchison Meteorite.

4. **A Conference in Sri Lanka**

With a confluence of observations and data from many directions pointing to a cosmic origin of life a rare opportunity came along for me to organise an international conference between 1$^{st}$ and 11$^{th}$ December 1982 in Colombo, Sri Lanka. The conference had the title "Fundamental Studies and the Future of Science" and was convened by the Institute of Fundamental Studies (IFS) and the office of His Excellency the President of Sri Lanka. The participants included Fred Hoyle and several key players in the unfolding drama of cosmic life. Fred Hoyle gave the first talk entitled "From Virus to Man" and Hans Pflug followed this with a breath-taking presentation of his work on meteorite microfossils. His Excellency President Junius Jayawardene (incidentally a fan of Fred Hoyle) sat speechless in the front row of the audience.

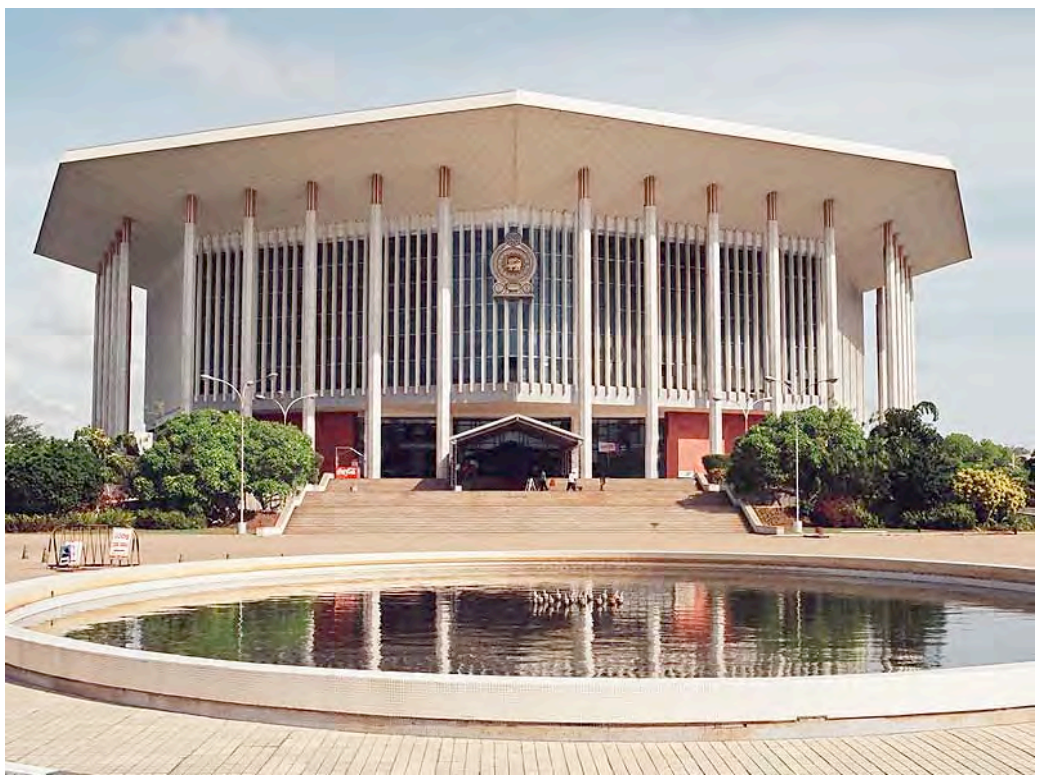

Fig. 3. The Bandaranaike Memorial Conference Hall in Colombo. Venue for the conference in December 1982 presided over by His Excellency the President of Sri Lanka, and including amongst others: Sir Fred Hoyle, Hans Pflug, Bart Nagy, Gustaf Arrhenius, Cyril Ponnamperuma, Phil Solomon, Sir Arnold Wolfendale and Chandra Wickramasinghe.

Just as one felt that the game was up for Earth-centred life, Gustaf Arrhenius, the grandson of Svante Arrhenius, slated Pflug mercilessly for reviving the ignominous microfossil saga of the 1960's. One is enjoined, he argued, not to pay any attention to morphology. All that was permitted to take place in meteorite formation processes are catalysis, chemistry and mineralogy. What Hans Pflug showed us in Colombo were artifacts! Life look-alikes, not life!

After much heated arguments about microfossils, Sri Lankan-born Cyril Ponnamperuma's lecture trawled through extensions of the Miller-Urey synthesis adding little to the debate. The question that reverberated through the lofty spaces of the Chinese-built BMICH auditorium was stark and simple: "Was the evidence we saw for extraterrestrial life real or illusory?" Heated arguments continued late into the night in the hotel bar of the Oberoi – the luxury hotel where the participants stayed. It was amusing to hear pontification by Gustav Arrhenius denouncing his grandfather Svante for ever raising the spectre of panspermia. It was obvious that there could be no consensus. Bartholamew Nagy, who was due to speak on the next day, was seen scampering between rival cliques like a rabbit scared out of its wits.

Nagy had been billed to give a talk in support of Pflug, but in the event he delivered a talk with the unimposing title: "Search for potential biochemical fossils less than $3.5 \times 10^9$ years old", concentrating on the safe option of microfossils in terrestrial sediments. The irony was that whilst microfossils in terrestrial sedimentary rocks were considered valid observations, similar or even identical structures in extraterrestrial rocks were necessarily all contaminants!

Atmospheric physicist E.K. Bigg next showed images of the very earliest microscopic particles collected at great heights in the stratosphere, some of which may be identified with microbial particles entering the Earth from outside. Bigg was understandably guarded in his conclusions, however, so he did not attract the hostility that Hoyle, Pflug or Nagy had received. It is ironical that weak evidence often goes unchallenged; the stronger the evidence that challenges a reigning paradigm, the more vehement the protestations and rebuttals tend to be.

Phil Solomon (SUNY), the pioneer of interstellar CO observations, summed up the session with a brilliant talk on "Molecular Clouds in the Galaxy", which after all was where the modern revival of panspermia began in the 1970's.

Throughout this historic meeting, whether in hotel bars, or as we toured the magnificent beaches and historical monuments of Sri Lanka, the arguments about alien life never ceased. With all the relevant people assembled at the same place at the same time it felt as though a momentous paradigm shift was in the air. But that was not to be. Conceding anything at all seemed difficult. There was a great hesitancy to admit anything at all that might "rock the boat" – a boat that now seemed to be sailing in stormy tropical waters of the Indian Ocean, at any rate from the vantage point of conservative science.

Sir Arnold Wolfendale, former Astronomer Royal, had friends in Sri Lanka and felt very much at home there, but not so with the portents of a paradigm shift that appeared ominously on the horizon. He proposed that the matters discussed in Colombo be *properly* resolved at a discussion meeting of the Royal Astronomical Society, a meeting that he as President of the RAS was able to convene. Thus ended a conference in tropical Sri Lanka with what seemed to be a disappointingly indecisive result. Despite the weight of facts and evidence presented, and the good humour that prevailed, few admitted to have changed their mind. As Julius Caesar said "People willingly believe what they want to believe..." (*The Gallic War*, 3.18)

5. **Meeting of the Royal Astronomical Society**

As promised by Wolfendale a discussion meeting of the Royal Astronomical Society with the title "Are interstellar grains bacteria?" took place at the Scientific Societies Lecture Theatre in Savile Row London on 11th November 1983. Fred Hoyle and I presented arguments for our thesis of cosmic life and several others including Nobel Laureate Harry Kroto followed on with what amounted to no more than polemical retorts. Phil Solomon summarised the turbulent proceedings thus:

"A strong case for identification of grains with bacteria has been made and nothing yet presented appears to destroy the case. The issue is one which in spite of its dramatic implications can be settled in the near future by further observations....."

Summaries of the papers presented at this meeting are published in *The Observatory*, Vol 104, No.1060, pp129-139, June 1984.

6. **Further Developments**

Further studies and more extensive observations of interstellar dust, comets, cometary debris and meteorites have added to the evidence in quick succession (J.Wickramasinghe et al, 2010; Smith et al, 2007). Astronomical observations deploying new telescopes and instruments yielded spectroscopic data that confirms the dominance of complex organic molecules similar to biological degradation products in interstellar clouds and even in distant galaxies at high redshifts (Wickramasinghe et al 2004; Wickramasinghe, 2010; Gibson and Wickramasinghe, 2010). Hydrogravitational dynamics HGD theory supports cometary panspermia and the early formation of complex life, in 705 F hot water oceans of primordial hydrogen planets that formed the first stars and chemicals (Gibson, Schild and Wickramasinghe 2010; Gibson, Wickramasinghe and Schild, 2011). A cosmological primordial soup (Miller and Urey 1959) existed on $10^{80}$ merging primordial gas planets and their comets, all tracking progress in organic chemistry among the planets at time 2 Myr when the density of the cosmos was larger than the density at present by a factor of $10^8$. Thirty million frozen primordial gas planets per star in protoglobularstarclusters make up the dark matter of galaxies according to HGD cosmology.

Extensive studies of comets after the 1986 perihelion passage of Comet Halley and the *Giotto Mission* showed to the consternation of most astronomers that cometary dust did indeed contain a large fraction of complex organics with spectra resembling biomaterial. Whipple's dirty snowball model of comets was seriously challenged. Cometary dust collections in the *Stardust Mission* to Comet Wildt 2 also confirmed the presence of either degradation products of life or the building blocks of life, in the destruction trails left by high-speed dust particles as they were slowed in blocks of aerogel. Unfortunately, however, because the aerogel was not sterilized, no biological inferences could be drawn. The Deep Impact Mission to comet Tempel 1 in 2004 once again confirmed the organic composition of comets but also showed evidence of clay minerals, consistent with the presence of liquid water at an early stage in the comet's history (A'Hearn et al, 2005; Lisse et al, 2006). This is important because it would show that comets could provide habitats for the replication of cosmic bacteria – thus making these celestial bodies viable homes for life.

The 2001-2009 ISRO-Cardiff studies of dust recovered using balloons flown to heights of 41km in the stratosphere showed not only the presence of viable cultures of UV resistant microorganisms, but also evidence for larger 15-20 micron-sized clumps of dormant microorganisms (Narlikar et al, 2003; Wainwright et al, 2003). The latter were detected with the use of fluorescent dyes showing the presence of viable but not culturable microorganisms. The possibility of terrestrial contamination was virtually excluded, the height of 41km being too high for lofting large clumps of terrestrial material from the surface; and so we were led to conclude that here is evidence of a continuing incidence of microorganisms from space. From our collection statistics we estimated a daily input of viable biomaterial of about 0.1 tonne averaged over the whole Earth.

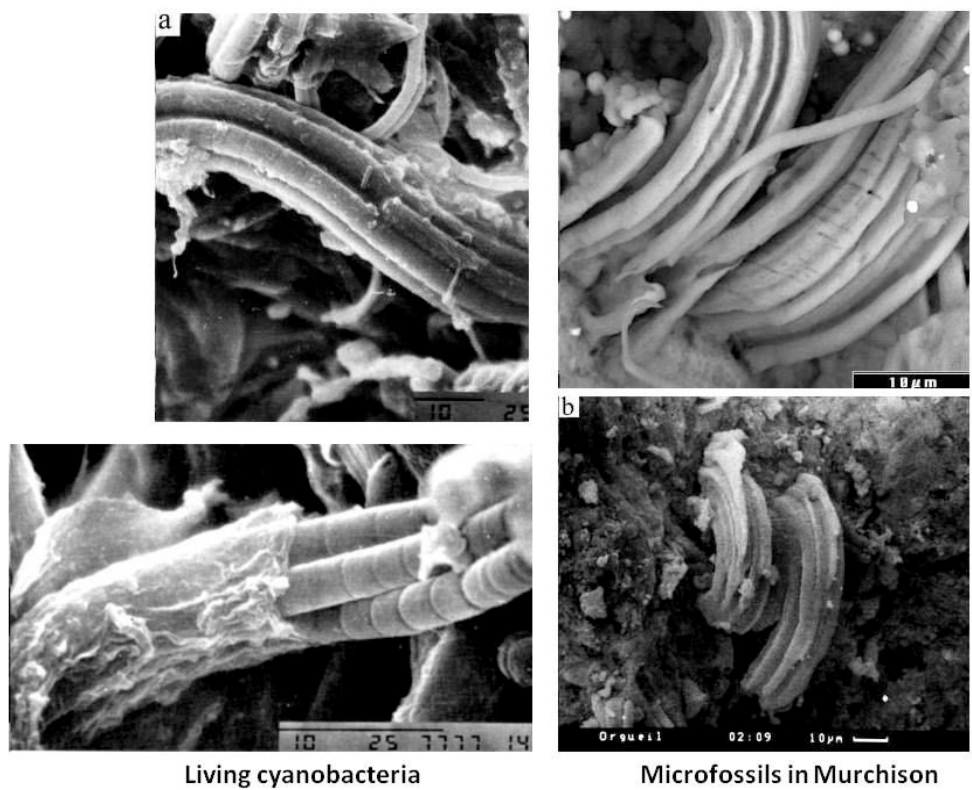

Living cyanobacteria      Microfossils in Murchison

Fig. 4 Hoover's Murchison microfossils of cyanobacteria compared with modern cyanobacteria (Hoover, 2005)

It is against this backdrop that one should approach Richard B Hoover's recent re-examination of Murchison meteorite microfossils (Hoover, 2005, 2011). (See for instance Fig.4). Using state of the art technology Hoover concludes that microbial fossils unambiguously exist in great profusion. The furore that greeted this new publication, with vocal condemnation from Science journals and from NASA chiefs, shows that earlier tactics of rejection by silence have now been replaced by strident ranting and even personal insults. Had we lived in the Middle Ages there is no doubt that Richard B Hoover, and possibly Fred Hoyle, Pflug, and I too, would have come to a bad end – suffering the fate of Giordano Bruno in 1600!

The handicap facing 21$^{st}$ century science is an excess of specialisation. Although some extent of specialisation is predicated by the huge quantity of information within each separate discipline, the disadvantage is that cross-disciplinary developments are discouraged. This, I believe, is one factor at least that has impeded the acceptance of the cosmic theories of life. To astronomers the association of bacteria with interstellar grains would appear understandably strange; and to biologists the intrusion of astronomy into their discipline would be equally repugnant. But the Universe encompasses everything, and is no respecter of any man-made boundaries between disciplines.

In the year 2011 all the relevant facts appear to converge in favour of the cosmic origins of life. Resisting the facts and imposing censorship would in the long run turn out to be futile. The Universe will always have its last say.

(Note: The proceedings of the 1982 conference in Sri Lanka is published in Wickramasinghe, 1984 (ed) detailed in a reference below.)